\documentclass{article}

\usepackage[
backend=biber,
style=ieee,
]{biblatex}

\title{Bibliography management: \texttt{biblatex} package}
\author{Share\LaTeX}
\date{ }

\usepackage{PRIMEarxiv}

\usepackage[utf8]{inputenc} 
\usepackage[T1]{fontenc}    
\usepackage{hyperref}       
\usepackage{url}            
\usepackage{booktabs}       
\usepackage{amsfonts}       
\usepackage{nicefrac}       
\usepackage{microtype}      
\usepackage{lipsum}
\usepackage{fancyhdr}       
\usepackage{graphicx}       
\graphicspath{{media/}}     

\title{Setting the rhythm scene: deep learning-based drum loop generation from arbitrary language cues
}

\author{
  Ignacio J. Tripodi \\
  Independent Researcher \\
  \texttt{ignacio.tripodi@colorado.edu} \\
}

\begin{document}
\maketitle

\begin{abstract}
Generative artificial intelligence models can be a valuable aid to music composition and live performance, both to aid the professional musician and to help democratize the music creation process for hobbyists. Here we present a novel method that, given an English word or phrase, generates 2 compasses of a 4-piece drum pattern that embodies the ``mood'' of the given language cue, or that could be used for an audiovisual scene described by the language cue. We envision this tool as composition aid for electronic music and audiovisual soundtrack production, or an improvisation tool for live performance. In order to produce the training samples for this model, besides manual annotation of the ``scene'' or ``mood'' terms, we have designed a novel method to extract the \textit{consensus} drum track of any song. This consists of a 2-bar, 4-piece drum pattern that represents the main percussive motif of a song, which could be imported into any music loop device or live looping software. These two key components (drum pattern generation from a generalizable input, and consensus percussion extraction) present a novel approach to computer-aided composition and provide a stepping stone for more comprehensive rhythm generation.
\end{abstract}


\section{Introduction}
The task of computer-based percussion track generation has mainly been focused on genre-specific rhythm, such as rock, pop, or jazz. Recent efforts have achieved impressive performance in this realm\cite{roberts_magenta_2019}. Our scope is rather different than the genre-focused approaches, however: we aim to take any word or short phrase as input, and use this cue to produce a 4-piece percussive track that embodies the ``mood'' of such input text. 

To train such a model, we needed annotated music samples with both words based on the music alone (not the lyrics, if featured), and a drum track in a machine-readable format that would summarize the main percussive motif of this song and could be used to loop over. The word annotation task unfortunately required a manual, time-consuming human curation effort, as to our knowledge no such corpus exists or is publicly available. The drum patterns to use as training, however, would have taken a significantly longer time to create manually, making it impractical for the number of training samples required. The following sections describe how we have automated the extraction of a consensus 2-bar loop drum pattern from any given song, directly from the digital audio file, then we manually annotated language tokens describing the mood/scene of the song's rhythmic pattern, and trained a model to produce new percussive loop sequences based on any arbitrary language cue.

\section{Methods}
\label{sec:headings}

The encoding task to translate from words to drum loop info has been approached with a deep multivariate regression network. The input consisted of word embeddings, which are vectors of real numbers representing the words' semantic meaning. The predicted output consists of a 129-dimension vector, that includes the suggested tempo and four percussion instrument representation patterns along 2 bars (each size 32).

The dataset required to train this model consisted of one or more keywords to represent the ``mood'' or ``situation/scene'' of the music alone (ignoring the content of the lyrics, if those where available), as well as a 2-bar drum loop that represented the most common rhythm of that song. Since no available corpus of consensus drum loops and mood keywords exists, we had to create one.


\subsection{Language token annotation}

A total of 328 songs from diverse styles have been manually tagged with words that come to mind listening to the song, as well as descriptions of audiovisual scenes the song would be a good fit for. For example, a song that could normally be labeled as ``hard rock'' was tagged with language tokens such as ``black leather'', ``cigarette smoke'' or ``headbanging''; similarly, a song that could be usually tagged as ``trance'' or ``breakbeat'' has been labeled as ``tension'' and ``persecution''. The word annotations for the songs often considered a visual ``scene'' for which the song would be a suitable soundtrack, i.e., ``car chase''.  This portion of the corpus generation had to be performed manually for every song. Word embeddings were employed to computationally represent the mood terms. We have used a pre-trained BERT\cite{devlin_bert_2019} model to produce 768-dimension input vectors to our model by averaging the embeddings for all terms assigned to a song. Table \ref{table:mood_label_examples} illustrates a handful of manual annotation examples.

\begin{table}
 \caption{Song annotation examples}
  \centering
  \begin{tabular}{lll}
    \toprule
    Artist     & Song     & Word/phrase annotations \\
    \midrule
    Iron \& Wine & Peng!                & ``peaceful'', ``soft''     \\
    Björk       & Hunter                & ``marching'', ``approaching something''      \\
    Oasis       & Songbird              & ``happy'', ``cheerful'', ``sweet''  \\
    Devo        & Through Being Cool    & ``get up and go'', ``activate''  \\
    The Prodigy & Firestarter           & ``speed'', ``aggression'', ``tension''  \\
    Thom Yorke  & Truth Ray             & ``downbeat'', ``grieving'', ``weary'' \\
    \bottomrule
  \end{tabular}
  \label{table:mood_label_examples}
\end{table}

\subsection{Consensus drum loop extraction}

Another novel aspect of this work is a method to extract a 2-bar consensus drum beat for any song, directly from the audio files in MP3 format. Given that manually producing this ``consensus'' drum loop to illustrate the common song rhythm would have been exceedingly time-consuming, we have devised a model that extracts such drum pattern programmatically. We have chosen 2 bars (2 compasses) as the output length as it's a good fit for common looping purposes and music sequencer hardware. We focused on generating 4-piece percussion patterns, in line with most common drum machine capabilities. This has worked well with most songs, while in a few cases the percussive sounds were too simple or too hard to tell apart (see method below), which resulted in a consensus pattern of fewer than 4 percussive instruments.

All songs chosen to train, test and validate the model include some form of percussion, either acoustically- or electronically-generated. Using the librosa\cite{mcfee_librosalibrosa_2022} Python library, we have performed a Short-Time Fourier Transform (a set of discrete Fourier transforms over short overlapping windows) to analyze the audio data in frequency space. We then performed median-filtering harmonic percussive source separation\cite{fitzgerald_harmonicpercussive_2010,driedger_extending_2014} using a kernel size of 75x75, determined empirically after a grid search of kernel dimensions (all pairwise combinations of 50, 75, and 100, including squares) . Rather than applying librosa's tempo detection method directly to the song, we have found that (for our training dataset) a more accurate estimation resulted from generating a wavefile of clicks at the strongest beat time points. Utilizing librosa's onset detection method, we gathered all time points at which onsets at the 98th percentile of envelope strength were detected (usually snare drum hits), thus keeping the strongest beats. The wavefile of clicks at these time points was then fed as input to librosa's tempo estimation method, generally resulting in more accurate tempo values.

\begin{figure}
  \centering
  \includegraphics[width=0.6\textwidth]{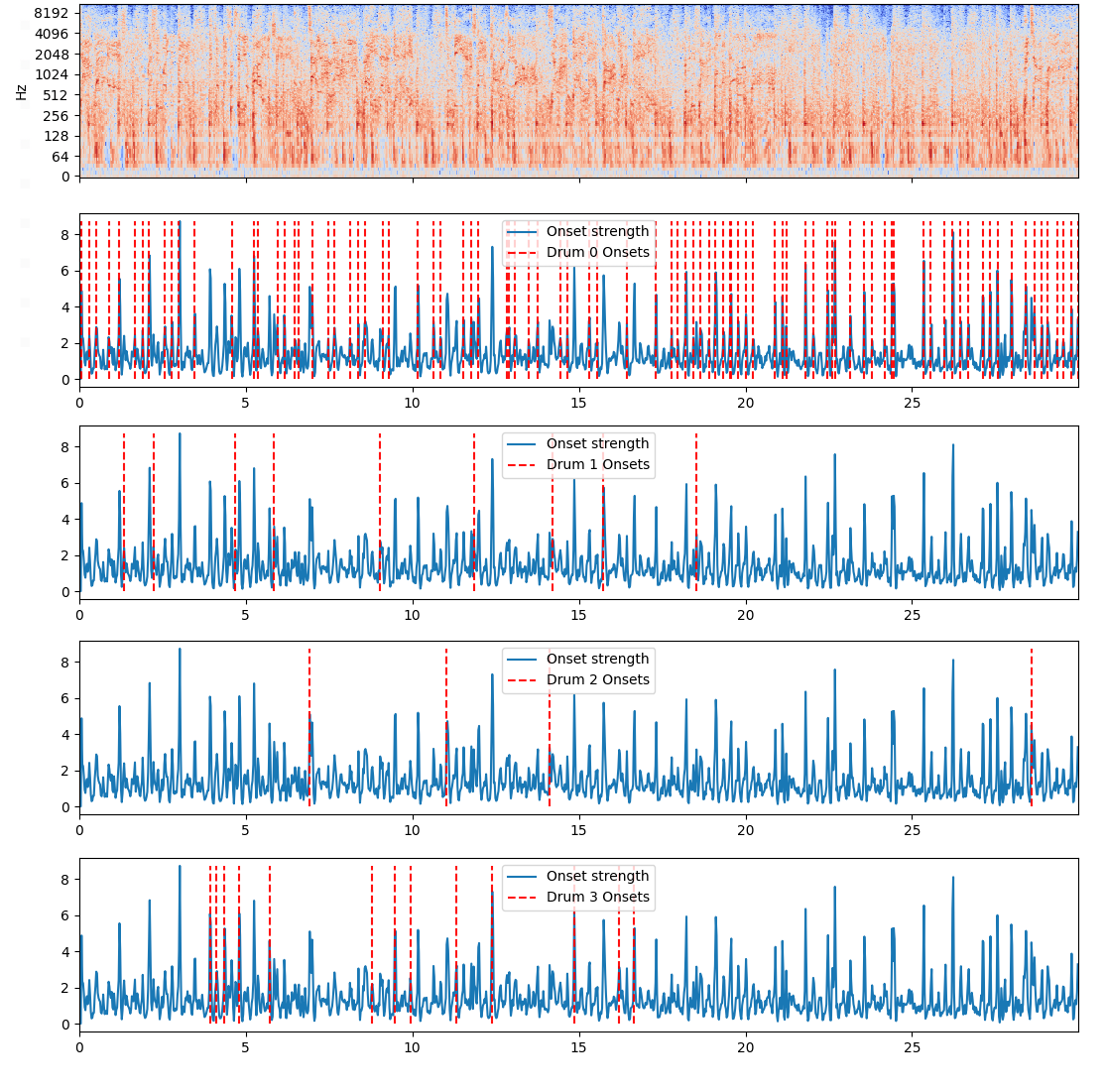}
  \caption{Percussive onset detection and clustering: Percussive onsets were detected using the Librosa library and clustered based on their energy along the frequency spectrum. 
  }
  \label{fig:four_freq_spectrum_plots_with_onsets}
\end{figure}

To prevent any noise from special song introductions of varying lengths, we have started sampling the music files at 60 seconds, for a duration of 60 more seconds (songs shorter than 2 minutes were not considered). Based on the detected tempo and starting time of the song sampling process, we were able to accurately detect the first ``beat one'' percussive onset as reference for the rest. For simplicity, we have assumed that all songs follow a 4/4 rhythm and used a resolution of 16th notes. Each detected percussive onset's time along the 60-second sample was fitted to its nearest discrete 16th note bin, which we have considered sufficiently precise for most drum machine capabilities. To separate the 4 distinct types of drum instruments, we performed k-means unsupervised clustering using the frequency domain vector for each beat onset, representing the different energy values at each frequency range. This resulted in onset times for up to 4 different 32-sized patterns (2 bars of size 16 each) to fit most common drum machine hardware (Fig. \ref{fig:four_freq_spectrum_plots_with_onsets}).

\begin{figure}[h!]
  \centering
  \includegraphics[width=0.6\textwidth]{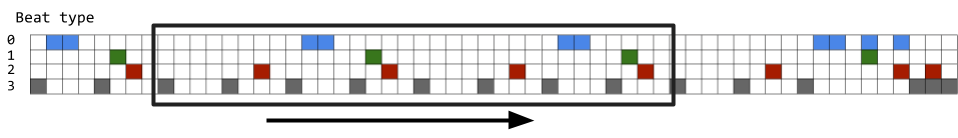}
  \caption{Sliding window approach: after onset detection, we scan the entire sample to look for the most commonly occurring 2-bar pattern across the four detected percussive tracks. }
  \label{fig:sliding_window}
\end{figure}

A sliding window approach was then taken to scan thru the song sample and look for the most commonly occurring 32-sized sequence across all 4 percussive instruments (Fig. \ref{fig:sliding_window}). In the case of a tie, the sequence featuring beats from the greatest number of percussive instruments was favored. The detected tempo and the 4 binary vectors of size 32 each, denoting where there is a beat for each of the 4 percussive instruments, constitutes the detected 129-dimension (1 + 32 + 32 + 32 + 32) consensus rhythm output vector. This method also suggests which of the 4 clustered percussive tracks corresponds to a low frequency beat (e.g., kick drum), a strong percussive beat (e.g., snare drum), a short decay cymbal (e.g., hi-hat) and another percussive sound among mid frequency range (e.g., tom, conga, cowbell). The percussive instrument suggestion for each track is based on the median envelope strength (provided with the beat onset calculation) for all beats in that track. The order of increasing median envelope strength that has worked best for the majority of songs was: snare drum, kick drum, other percussion, hi-hat.

\subsection{Neural network architecture}

\begin{figure}[h!]
  \centering
  \includegraphics[width=0.6\textwidth]{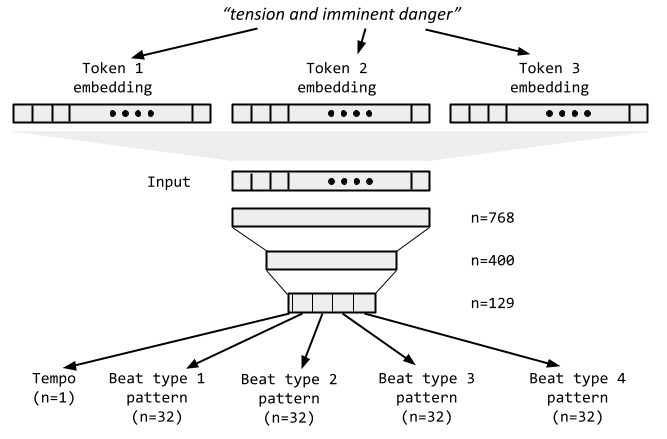}
  \caption{Neural network design.}
  \label{fig:nn_design}
\end{figure}

The neural network trained to predict a drum loop from a language token (or set of tokens) consisted of a fully-connected network with an input layer of size 768 (matching the word embedding dimensions), a dense layer of size 400, and an output layer of size 129 (Fig. \ref{fig:nn_design}). A deep regression model was trained using 328 songs for which the language tokens have been manually annotated, and the consensus drum loop extracted using the method described above. For the input layer we used rectified linear unit (ReLU) activation, and a a uniform variance scaler initializer\cite{he_delving_2015} for the mid and output layers. A custom activation function was used for the output layer, setting to zero any values below the 75\textsuperscript{th} percentile. This multivariate regression model was trained using a batch size of 5 and the Huber loss function, which has performed better than the mean absolute error (MAE) and mean squared error (MSE) loss. Using the Adam optimizer, we evaluated the model's performance using 10-fold cross validation over 3 repeat runs.

During development, the Online Sequencer\cite{noauthor_online_nodate} web application has been used to qualitatively evaluate the output for rhythm usefulness and expressiveness. The suggested order of beat types has been mapped to Online Sequencer's ``Drum Kit'' instruments library as follows: ``High Tom 1'' (D4), ``Kick Drum 1'' (C3), ``Mute Hi Conga'' (D5), and ``Closed Hi-Hat'' (F\#3), respectively.

\section{Results}\label{sec:headings}

After finding the optimal set of hyperparameters, the network has been trained with all annotated samples and initially tested with a list of terms for a qualitative evaluation of the output. Given a word or phrase, the averaged word embeddings were calculated using the pre-trained BERT model to use as input to the neural network. From the predicted output vector, the first value (the suggested tempo) was left as-is, and for the remaining 128 values in the multivariate regression output we considered those above the 75 percentile (3rd quartile) to represent a beat in that position, otherwise a silence was assumed. 

The current implementation outputs the suggested tempo for the given keywords, and a text diagram indicating the position of a 16th note beat along the 2 compasses for each beat type. Using an ``X'' to indicate where there is a beat, a ``-'' to indicate a silence and ``|'' to separate a compass, it provides a simple visualization to program any available drum machine hardware. It also produces a string that can be copied and pasted into the Online Sequencer web application, to easily listen to the generated loop. For example, given an input of ``aggressive attack'', the formatted output is available below:

\begin{verbatim}
Suggested tempo: 121
	|-------------X--|----------X-----|
	|--X-X-X--X-X----|X-----------X---|
	|--X-X---X------X|X--X-----------X|
	|-----X--XXXXXX--|X--X-XXXX-X--XX-|

Online Sequencer:319887:13 F#3 1 2;26 F#3 1 2;2 D5 1 2;4 D5 1 2;6 D5 1 2;9 D5 1 2;
11 D5 1 2;16 D5 1 2;28 D5 1 2;2 C3 1 2;4 C3 1 2;8 C3 1 2;15 C3 1 2;16 C3 1 2;19 C3 1 2;
31 C3 1 2;5 D4 1 2; 8 D4 1 2;9 D4 1 2;10 D4 1 2;11 D4 1 2;12 D4 1 2;13 D4 1 2;16 D4 1 2;
19 D4 1 2;21 D4 1 2; 22 D4 1 2;23 D4 1 2;24 D4 1 2;26 D4 1 2;29 D4 1 2;30 D4 1 2;:
\end{verbatim}

\begin{figure}[h!]
  \centering
  \includegraphics[width=0.7\textwidth]{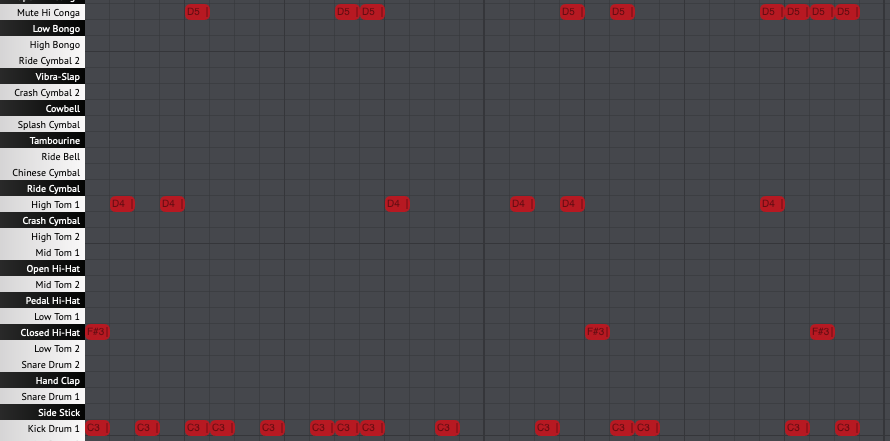}
  \caption{Drum pattern generated by the input phrase ``tension and imminent danger'' (predicted tempo of 129bpm), loaded in the Online Sequencer web app. An audio clip of three loop iterations using the default Drum Kit instrument samples can be accessed at \url{https://zenodo.org/record/7067265}.}
  \label{fig:tension_and_imminent_danger_drums}
\end{figure}

\begin{figure}[h!]
  \centering
  \includegraphics[width=0.7\textwidth]{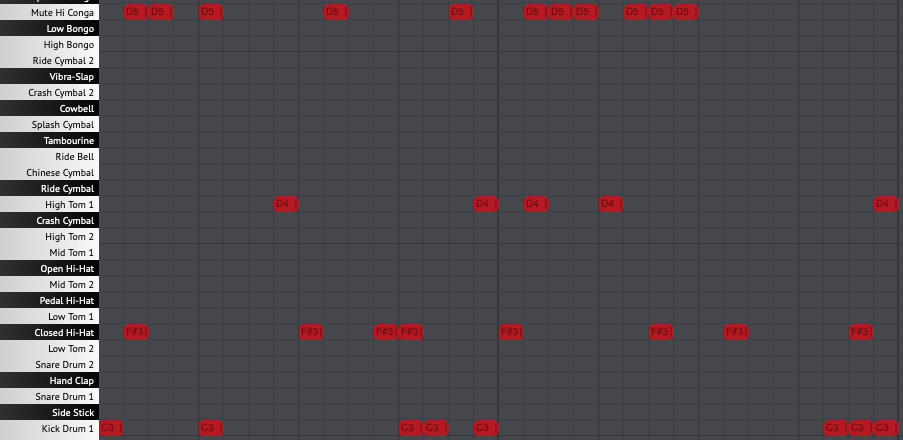}
  \caption{Drum pattern generated by the input word ``depressed'' (predicted tempo of 77bpm), loaded in the Online Sequencer web app. An audio clip of three loop iterations using the default Drum Kit instrument samples can be accessed at \url{https://zenodo.org/record/7067265}.}
  \label{fig:depressed_drums}
\end{figure}

The Online Sequencer proved to be a practical and time-efficient way to evaluate the generated drum pattern for any given word or phrase. Figures \ref{fig:tension_and_imminent_danger_drums} and \ref{fig:depressed_drums} illustrate how the drum pattern is visualized for two example inputs, ``tension and imminent danger'' and ``depressed'', respectively. Audio files of synthesized instruments for 3 loop iterations of these drum patterns can be obtained at \url{https://zenodo.org/record/7067265}.

The pre-trained model with the manual annotations and Python code to produce a 4-sample, 2-bar percussion sequence along with a suggested tempo can be accessed at \url{https://github.com/ignaciot/words2drums}.


%

\section{Discussion}\label{sec:headings}

We presented two complementary novel methods, one in computational music information extraction and another in synthetic rhythm generation. On one hand, we demonstrated how we can produce a consensus 2-bar, 4-sample drum pattern for any given music file (currently MP3 or wave file, but easily expandable to other formats). On the other hand, we used the extracted consensus drum patterns with manually annotated language tokens to train a model that produces a new drum pattern based on any given word or phrase, that would capture the ``mood'' or ``scene'' for those words. While keeping into account that the output validation is inevitably qualitative, the authors have found this model to produce interesting and useful drum patterns that could be used to easily program a drum machine as a starting point, aiding the creative process. We also acknowledge a larger sample size is highly desirable to extend this proof of concept into a more robust application.

\subsection{Future work}

It would be interesting to expand the consensus drum loop extraction capabilities to other rhythms than 4/4. Other output formats can also be considered for the synthesized drum pattern, to increase this tool's accessibility. We also acknowledge the limitations of expressiveness that a percussion pattern can convey, making the language annotations highly subjective; however, we consider this to be the first step towards scene-based rhythm generation; besides the drum pattern and tempo, a natural extension to this work would be the generation of instrument configurations (such as synthesizer envelope parameters, sample choice, etc.) that go along with the drumming pattern. A more interactive use of this model would also be desirable, for example producing a MIDI output stream that can be directly sent to a sequencer, rather than having to manually enter the drum pattern.

\section*{Acknowledgments}
We would like to thank Sam Way for his feedback and encouragement, and Carlos Falanga, Behzad Haki, and Dr. Sergi Jordà for their valuable feedback of the working prototype.


\medskip

\printbibliography

\end{document}